\documentstyle[12pt]{article}

\begin{document}

\title{Statistical approach of the modulational instability of the discrete 
self-trapping equation}

\author{Anca Vi\c sinescu, D. Grecu\\
{\small \it  Department of Theoretical Physics} \\ 
{\small \it National Institute for Physics and Nuclear Engineering}\\
{\small \it "Horia Hulubei"}\\
{\small \it P.O.Box MG-6, M\v agurele, Bucharest, Romania}\\
{\small \it e-mail: avisin@theor1.theory.nipne.ro}\\
{\small \it ~~~~~~~~~~dgrecu@theor1.theory.nipne.ro}}
\date{}
\maketitle

\begin{abstract}

The discrete self-trapping equation (DST) represents an useful model 
for several properties of one-dimensional nonlinear molecular crystals. 
The modulational instability of DST equation is discussed from a 
statistical point of view, considering the oscillator amplitude as a 
random variable. A kinetic equation for the two-point correlation 
function is written down, and its linear stability is studied. Both a 
Gaussian and a Lorentzian form for the initial unperturbed wave 
spectrum are discussed. Comparison with the continuum limit (NLS 
equation) is done.

\end{abstract}

\vskip 1cm
\section{Introduction}
\vskip .5cm

The discrete self-trapping (DST) equation
\begin{equation}\label{1}
i{da_{n}\over dt}-\omega_{0}a_{n}+\lambda (a_{n+1}+a_{n-1})+\mu \vert 
a_{n}\vert^{2}a_{n}=0
\end{equation}
is a typical equation for a system of harmonically coupled nonlinear 
oscillations \cite{1}, \cite{2} relevant for several physical problems. 
We mention here only Davydov's model of energy transport in 
$\alpha$-helix structures in proteins \cite{3}, \cite{4}, \cite{2}, 
where (\ref{1}) appears as a certain approximation of the model. In 
(\ref{1}) $a_{n}$ is the complex classical dimensionless amplitude of 
the oscillator of frequency $\omega_{0}$ in the $n$-th molecule, and 
$\lambda , \mu $ (of dimension of frequency) are the coupling constants 
between nearest neighbour oscillators and the one-site nonlinearity 
respectively. It is well known that depending upon of the parameters 
and the chosen initial condition  the equation (\ref{1}) can lead 
either to self-trapping (i.e. local modes or solitons), or to chaos, or 
to a mixture of the above two behaviours \cite{1}, \cite{2}, \cite{5}. 
Instead of (\ref{1}) we shall consider the equation
\begin{equation}\label{2}
i{da_{n}\over dt}+\lambda (a_{n+1}+a_{n-1})+\mu \vert 
a_{n}\vert^{2}a_{n}=0 
\end{equation}
which is obtained if $a_{n}\rightarrow a_{n}e^{-i\omega_{0}t}.$ This 
equation admits plane wave solutions with constant amplitude
$$
a_{n}=ae^{i(kn-\omega t)} \nonumber
$$
(the lattice constant is taken equal with unity) but with an amplitude 
depending dispersion relation
$$
\omega (k)=-2\lambda\cos k -\mu \vert a\vert^{2} \nonumber
$$
This is a Stokes wave solution and it is well known to be unstable 
at small modulation of the amplitude (Benjamin-Feir or modulational 
instability) \cite{6}-\cite{8}. The aim of this note is to study the 
modulational instability of equation (\ref{2}) from a statistical point 
of view, considering $a_{n}$ as a random variable. In doing this we 
shall follow the procedure used by several authors to discuss the 
effects of randomness on the stability of weakly nonlinear waves, 
especially in hydrodynamics \cite{9}, \cite{10}.

In the next section a kinetic equation for a two-point correlation 
function will be obtained. Using a Wigner-Moyal transform 
the equation is written in a mixed configuration-wave vector space. The 
linear stability around a homogeneous basic solution is discussed in 
section 3. An integral stability equation is derived, very similar 
with the dispersion relation of the linearized Vlasov equation in 
ionized plasmas. Two forms for the spectrum of the initial unperturbed 
condition will be considered, namely a Gaussian and a Lorentzian form 
and in the limit of vanishingly small bands widths the increment of the 
modulational instability is calculated. Comparison with the continuum 
limit, when (\ref{1}) transforms into the nonlinear Schr\"odinger 
equation is done. Few concluding remarks are also presented.

\vskip .5cm

\section{Kinetic equation for two-point correlation function}

\vskip .5cm

Introducing the displacement operator by $a_{n\pm 1}=e^{{\partial \over 
\partial n}}a_{n}$ the equation (\ref{2}) becomes
\begin{equation}\label{3}
i{\partial a_{n}\over \partial t}+2\lambda \cosh {\partial \over 
\partial n}a_{n}+\mu \vert a_{n}\vert^{2}a_{n}=0.
\end{equation}
In order to find a kinetic equation we write (\ref{3}) for $n=n_{1},$ 
multiply it by $a_{n_{2}}^{*},$ add it to the complex conjugated of 
(\ref{3}) for $n=n_{2}$ multiplied by $a_{n_{1}}$ and finally take an 
ensemble average. One obtains
$$
i{\partial \over \partial t}<a_{n_{1}}a^{*}_{n_{2}}>+2\lambda (\cosh 
{\partial \over \partial n_{1}}-\cosh {\partial \over \partial 
n_{2}})<a_{n_{1}}a^{*}_{n_{2}}>+ \nonumber
$$
\begin{equation}\label{4}
\mu 
(<a_{n_{1}}a^{*}_{n_{1}}a_{n_{1}}a^{*}_{n_{2}}>-<a_{n_{2}}a^{*}_{n_{2}}a
_{n_{1}}a^{*}_{n_{2}}>)=0 \nonumber
\end{equation}
which beside the two point correlation function $\rho 
(n_{1},n_{2},t)=<a_{n_{1}}(t)a^{*}_{n_{2}}(t)>$ contains also 
four-point correlation functions. If $a_{n}$ corresponds to a Gaussian 
process, and this property is retained during the evolution, a 
four-point correlation function factorizes exactly in products of 
two-point correlation functions \cite{11}
\begin{equation}\label{5}
<a_{n_{1}}a^{*}_{n_{1}}a_{n_{1}}a^{*}_{n_{2}}>=2<a_{n_{1}}a^{*}_{n_{2}}>
<a_{n_{1}}a^{*}_{n_{1}}>=2 \rho (n_{1},n_{2})\bar{a^{2}}(n_{1})
\end{equation}
where $\bar{a^{2}}(n)=<a_{n}a^{*}_{n}>$ is the ensemble average of the 
mean square amplitude. Although the factorization (\ref{5}) is true 
only for a Gaussian process we shall assume to be at least 
approximately valid also for processes slightly different from a 
Gaussian one, and it represents the main approximation of the present 
analysis.

It is convenient to use a Wigner-Moyal transform \cite{12}. One introduce the 
new variables
\begin{equation}\label{6}
M={n_{1}+n_{2}\over 2}, ~~~~m=n_{1}-n_{2}.
\end{equation}
Then the equation (\ref{4}) becomes
\begin{equation}\label{7}
i{\partial \rho \over \partial t}+4\lambda \sinh {1\over 2}{\partial 
\over \partial M} \sinh {\partial \rho \over \partial m}+2\mu 
(\bar{a^{2}}(M+{m\over 2})-\bar{a^{2}}(M-{m\over 2}))\rho =0
\end{equation}

We consider a chain of $N$ molecules and impose cyclic boundary 
conditions. The Fourier transform of the two-point correlation 
function is defined by
\begin{equation}\label{8}
F(k,M,t)=\sum_{m}e^{-ikm}\rho(M+{m\over 2}, M-{m\over 2},t)
\end{equation}
where $k$ takes values in the first Brillouin zone (BZ), $k\in (-\pi, 
\pi ).$ The inverse formula is 
\begin{equation}\label{9}
\rho(M+{m\over 2}, M-{m\over 2},t) ={1\over M}\sum^{ 
BZ}_{k}e^{ikm}F(k,M,t)={1\over 2\pi}\int^{\pi}_{-\pi}e^{ikm}F(k,M,t)dk.
\end{equation}
For $m=0$ one obtains
\begin{equation}\label{10}
\bar{a^{2}}(N,t)={1\over N}\sum^{BZ}_{k}F(k,M,t)={1\over 
2\pi}\int^{\pi}_{-\pi}F(k,M,t)dk.
\end{equation}
Now Fourier transforming equation (\ref{7}) we get 
$$
{\partial F\over \partial t}+4\lambda \sin k \sinh {1\over 2}{\partial 
\over \partial M}F + \nonumber
$$
\begin{equation}\label{11}
4\mu \sum^{\infty}_{j=1}{(-1)^{j\pm 1}\over 
(2j-1)!2^{2j-1}}\left({\partial^{2j-1} \over 
\partial M^{2j-1}}\bar{a^{2}}(M)
\right)\left({\partial^{2j-1} \over \partial 
k^{2j-1}}F(k,M)\right)=0
\end{equation}
which is the expected nonlinear evolution equation for $F(k,M,t)$ in a 
mixed configuration-wave number space $(M,k).$ Using the definition 
(\ref{8}) we see that $F(k,M,t)$ is a periodic function in the 
reciprocal space, $F(k+2\pi )=F(k).$

\vskip .5cm
\section{Stability analysis}
\vskip .5cm

As the unperturbed problem we shall consider a basic solution 
$F_{0}(k)$ independent of $M$ and $t.$ This is the random counterpart 
of the Stokes wave in a deterministic approach. A small perturbation 
around this homogeneous background is considered, namely
\begin{equation}\label{12}
F(k,M,t)=F_{0}(k)+\epsilon F_{1}(k,M,t)
\end{equation}
According to (\ref{10}) we have also
\begin{equation}\label{13}
\bar{a^{2}}(M,t)=\bar{a_{0}^{2}}+\epsilon \bar{a_{1}^{2}}(M,t)
\end{equation}
where
$$
\bar{a^{2}_0}={1\over 2\pi}\int_{-\pi}^{\pi}F_{0}(k)dk \nonumber
$$
\begin{equation}\label{14}
\bar{a^{2}_{1}}(M,t)={1\over 2\pi}\int^{\pi}_{-\pi}F_{1}(k,M,t)dk
\end{equation}
When (\ref{12}) is introduced into (\ref{11}), neglecting terms of 
order $\epsilon^{2},$ the following linear evolution equation for 
$F_{1}$ is obtained
$$
{\partial F_{1}\over \partial t}+4\lambda \sin k \sinh {1\over 2}{\partial 
 \over \partial M}F + \nonumber
$$
\begin{equation}\label{15}
4\mu \sum^{\infty}_{j=1}{(-1)^{j+1}\over 
(2j-1)!2^{2j-1}}~~{\partial^{2j-1} F_{0} \over 
\partial k^{2j-1}}~~{\partial^{2j-1}\bar{a_{1}^{2}}(M) \over \partial 
M^{2j-1}}=0
\end{equation}
Looking for a plane wave solution
$$
F_{1}(k,M,t)=f_{1}(k)e^{i(KM-\Omega t)} \nonumber
$$
after little algebra the following stability integral equation is found
\begin{equation}\label{16}
1+{\mu \over 4\pi \lambda \sin {K\over 
2}}\int^{\pi}_{-\pi}{F_{0}(k+{K\over 2})-F_{0}(k-{K\over 2})\over \sin 
k -{\Omega\over 4\lambda\sin {K\over 2}}}dk=0
\end{equation}
The modulational instability is related to $\Omega $ complex with a 
positive imaginary part, $Im \Omega >0.$ It is convenient to compare 
(\ref{16}) with the similar result for the continuum case of the 
nonlinear Schr\"odinger equation \cite{8}
\begin{equation}\label{17}
1+{\mu \over K\omega_{2}}\int^{\infty}_{-\infty}
{F_{0}(k+{K\over 2})-F_{0}(k-{K\over 2})\over 
k -{\Omega\over 2K\omega_{2}}}dk=0
\end{equation}
Although there are significant differences between the two expressions, 
when the width of the spectrum $F_{0}(k)$ is vanishingly 
small, the final results will look very similar.

\vskip .5cm

{\bf 3a. Gaussian spectrum}

\vskip .5cm

As a first example let us assume $F_{0}(k)$ to be a Gaussian function
\begin{equation}\label{18}
F_{0}(k)={\sqrt{2\pi}\over \sigma}\bar{a^{2}_{0}}e^{-{k^{2}\over 
2\sigma^{2}}}.
\end{equation}
This expression doesn't satisfy the periodicity condition 
 but for $\sigma $ vanishingly small the 
errors introduced are negligible. Also the relation (\ref{14}) is 
satisfied up to exponentially small terms.

It is convenient to introduce the new integration variable $t={1\over 
\sqrt{2}\sigma }(k\pm {K\over 2})$ and the notations
\begin{equation}\label{19}
z_{\pm}={1\over \sqrt{2}\sigma }\left({\Omega \over 2\lambda \sin K}\pm 
\tan {K\over 2}\right)
\end{equation}
$$
f_{\pm}(t)={1\over \sqrt{2}\sigma}\left(\sin \sqrt{2}\sigma t \pm \tan 
{K\over 2}(1-\cos \sqrt{2}\sigma t)\right). \nonumber
$$ 
Then (\ref{16}) becomes
\begin{equation}\label{20}
{\bar{a^{2}_{0}}\over \sqrt{2\pi}\sigma }~{\mu \over \lambda \sin 
K}\int^{{\pi\over \sqrt{2}\sigma}}_{-{\pi\over 
\sqrt{2}\sigma}}e^{-t^{2}}\left({1\over z_{+}-f_{+}}-{1\over 
z_{-}-f_{-}}\right)dt=1.
\end{equation}
In leading order in $\sigma $ the integral (\ref{20}) can be evaluated 
using the steepest descent method \cite{erde}. Denoting 
$G_{\pm}(t)=t^{2}+ln(z_{\pm}-f_{\pm}(t)), ~~~~t_{\pm}$ the zeros of the 
first derivatives ${dG_{\pm}(t)\over dt}=0, ~~~A_{\pm}={1\over 
2}{d^{2}G_{\pm}\over dt^{2}}$ for $t=t_{\pm},$ and extending the 
integration limits to infinity the integral is given by
\begin{equation}\label{21}
\sqrt{\pi }\left({1\over \sqrt{A_{+}}}e^{-G_{+}(t_{+})}-
{1\over \sqrt{A_{-}}}e^{-G_{-}(t_{-})}\right).
\end{equation}
In the limit $\sigma \ll 1$ we have approximatively \\
$ t_{\pm}\simeq {1\over 2z_{\pm}}=\sqrt{{\sigma \over 2}}{1\over 
{\Omega \over 2\lambda \sin K}\pm \tan {K\over 2}}, 
~~~~~~e^{-G_{\pm}(t_{\pm})}\simeq {1\over z_{\pm}} $ and $ A_{\pm}\simeq 
1.$\\
Then the integral becomes
$${-2\sqrt{2\pi}\sigma \tan{K\over 2}\over ({\Omega\over 2\lambda \sin 
K})^{2}-(\tan {K\over 2})^{2}}.$$
Considering 
$\Omega$ purely imaginary, $\Omega=i\Omega_{i},$ we finally get
\begin{equation}\label{22}
\Omega_{i}=4\lambda \sin {K\over 2}\sqrt{\bar{a^{2}_{0}}{\mu\over 
\lambda}-\sin^{2}{K\over 2}}
\end{equation}
and an instability is obtained $(\Omega_{i}>0)$  
if $\mu$ and $\lambda >0$ and if $\sin^{2}{K\over 
2}<\bar{a^{2}_{0}}~{\mu\over \lambda}.$

\vskip .5cm

{\bf 3b. Lorentzian spectrum}

\vskip .5cm

A simpler example is a Lorentzian form for $F_{0}(k)$
\begin{equation}\label{23}
F_{0}(k)=\bar{a_{0}^{2}}{p\sqrt{1+p^{2}}\over \sin{K^{2}\over 
2}+p^{2}}.
\end{equation}
It satisfies the periodicity condition and relation (\ref{14}).
The unperturbed two-point correlation function is easily calculated 
using (\ref{21}) in the definition relation (\ref{19}). Straightforward 
calculations give
\begin{equation}\label{24}
\rho_{0}(m)={\bar{a_{0}^{2}}\over [1+2p(\sqrt{1+p^{2}}+p)]^{m}}
\end{equation}
representing an exponentially decreasing law. For $p\ll 1$ we have 
$\rho_{0}(m)\simeq \bar{a^{2}_{0}}e^{-2pm}.$

In order to calculate the integral (\ref{16}) it is convenient to 
introduce the new integration variable $t=\tan{K\over 2}.$ Then the 
integral is over the whole real axis and can be done in the $t$-complex 
plane. In the new variable $F_{0}(k\pm{K\over 2})$ writes
\begin{equation}\label{25}
F_{0}(k\pm{K\over 2})\rightarrow 
{\bar{a_{0}^{2}}2p\sqrt{1+p^{2}}(1+t^{2})\over (1+\cos {K\over 
2}+2p^{2})t^{2}\pm 2(\sin{K\over 2})t+(1-\cos {K\over 2}+2p^{2})}
\end{equation}
having poles at
$$t^{+}_{1,2}= - a \pm ib~~~~~~~~~t^{-}_{1,2}=  a \pm ib $$
where
\begin{equation}\label{26}
a={\sin{K\over 2}\over 1+\cos {K\over 2}+2p^{2}}, ~~~~~ 
b={2p\sqrt{1+p^{2}}\over 1+\cos {K\over 2}+2p^{2}}.
\end{equation}
Considering $\Omega$ purely imaginary, $\Omega=i\Omega_{i}$ and 
denoting $z={\Omega_{i}\over 4\lambda \sin{K\over 2}}$ we have also
$$
{1\over {\Omega\over 4\lambda \sin {K\over 2}}-\sin k}\rightarrow -i 
{1+t^{2}\over zt^{2}+2it+z}
$$
having poles at
$$
t_{3}=i{\sqrt{1+z^{2}}-1\over z}~~~~~~t_{4}=-i{\sqrt{1+z^{2}}+1\over z}
$$
We shall consider $z$ as a small quantity and consequently $t_{4}\gg 1.$ 
Closing the contour in the lower complex half-plane $t$ its 
contribution can be neglected in the first order. Therefore we shall 
take into account only the poles $t^{(\pm)}_{2}$ and after 
straightforward calculations the relation (\ref{16}) becomes
\begin{equation}\label{27}
1={\mu \bar{a_{0}^{2}}\over \lambda \sin {K\over 2}}{MA+MX\over 
X^{2}+M^{2}}
\end{equation}
where
\begin{eqnarray}\label{28}
A=1+a^{2}-b^{2}, ~~~~~&&B=2ab \nonumber \\ 
X=zA+2b, ~~~~~&&M=2a-zB.
\end{eqnarray}
When $p\ll 1$ we approximate
\begin{equation}\label{29}
a\simeq {\sin{K\over 2}\over 1+\cos {K\over 2}}, ~~~~~b\simeq {2p\over 
1+\cos {K\over 2}}
\end{equation}
terms of order $p^{2}$ being neglected. Then (\ref{27}) can be 
considerably simplified and finally give us
\begin{equation}\label{30}
\Omega_{i}=4\lambda \sin {K\over 2}\left(\sqrt{{\mu\over 
\lambda}\bar{a^{2}_{0}}-\sin^{2}{K\over 2}}-2p{1+\cos {K\over 2}+\cos 
k\over 1+\cos {K\over 2}}\right).
\end{equation}
Modulational instability occurs for $\lambda$ and $\mu >0,$ 
$\sin^{2}{K\over 2}<{\mu\over \lambda\bar{a_{0}^{2}}}$ and if $p$ is 
smaller than a critical value. This result is similar with a previous 
one \cite{16} obtained with a simplified Lorentzian form for $F_{0}(k).$
Both results (\ref{22}) and (\ref{30}) can be compared with the similar 
results obtained in the NLS case \cite{8}.
\begin{eqnarray}\label{31}
&& \Omega^{(G)}_{i}=2K\omega_{2}\sqrt{{\mu\over 
\omega_{1}}\bar{a_{0}^{2}}-{K^{2}\over 4}} \nonumber \\
&& \Omega^{(L)}_{i}=2K\omega_{2}\left(\sqrt{{\mu\over 
\omega_{1}}\bar{a_{0}^{2}}-{K^{2}\over 4}}-p\right)
\end{eqnarray}
where the superscript $G/L$ refers to Gaussian/Lorentzian form of 
$F_{0}(k).$ It is easily seen that (\ref{31}) are obtained in a long 
wave limit $(K\ll 1).$ In the Lorentzian case both relations (\ref{30}) 
and (\ref{31}) show a behaviour similar with the well known phenomena 
of Landau damping in plasma physics \cite{13}, \cite{14} namely with 
increasing of $p$ the imaginary part $\Omega_{i}$ can become negative and 
no instability develops.

In conclusion a complete discrete discussion of the randomness effects 
on the MI of the self trapping equation was done and the discrete 
effects are easily seen in the final results, compared with the similar 
ones found for the NLS equation.

\vskip .5cm

{\small \it Helpful discussions with Dr. A.S. C\^arstea 
are kindly acknowledged. This research was supported under the 
contract 66, CERES Program, with the Ministry of Education and 
Research.}

\end{document}